\documentstyle[12pt,emlines]{article}\textwidth=165mm\textheight=245mm
\begin{document}

\title{Two-dimensional Cooling of Ion Beams in Storage Rings by Narrow
Broad-Band Laser Beams}\date{}
\author{E.G.~Bessonov, LPI AS, Moscow, Russia\\
K.-J.~Kim, ANL, APS, USA\\
F.~Willeke, DESY, Germany}
\maketitle

                       \begin{abstract}
A new scheme for the two-dimensional cooling of ion beams in storage
rings is suggested in which ions interact with a counterpropagating
broad-band laser beam. The interaction region in the direction of the
ion movement is much less then the wavelength of the ion betatron
oscillations. The laser beam in the orbit plane has sharp flat edge
directed to the ion beam and the width of the laser beam of the order
of the ion beam width. Laser beam radial position is being displaced
with some velocity at first from inside and then from outside to the
ion beam and decreases both betatron and synchrotron oscillations.
\end{abstract}

       \section{Introduction}

A series of the ion cooling methods are suggested to decrease the
emittances of charged particle beams. Among them: synchrotron radiation
damping \cite{kolom}, electron cooling \cite{budker}, stochastic
cooling \cite{vanderm}, one dimensional laser cooling \cite{wineland}.
Laser cooling is highly efficient in the longitudinal direction, but
difficult in the transverse direction unless a special coupling
mechanism is introduced \cite{sessler}, \cite{idea}. In papers
\cite{idea}-\cite{pac} the three-dimensional RIC method is proposed.
Nevertheless the quest for new cooling schemes remains vital for the
cooling of both not fully stripped and fully stripped high current
proton and more heavy ion beams. In this Letter, a new scheme for
two-dimensional cooling of ion beams in storage rings is suggested in
which photons in a counterpropagating narrow broad-band laser beam
interact with ions.

To explain the basic principle of the two-dimensional cooling of ion
beams in storage rings consider the process of change of the amplitude
of the betatron oscillations and the position of the instantaneous ion
orbit in the process of the energy loss of the ions. We will consider
the case when the RF accelerating system of the storage ring is
switched off.

Let us the instantaneous orbits of ions are distributed in a region
$\rho _0 \pm \sigma _{\rho \varepsilon}$ and the amplitudes of ion
oscillations are distributed in a region $\sigma _{b\varepsilon}$
relatively to the instantaneous orbits corresponding to the ion energy
$\varepsilon$, where $\rho _0 $ is the location of the middle
instantaneous orbit, $\sigma _{\rho \varepsilon} > 0$ mean-root square
deviation of the instantaneous orbits from the middle one which is
determined by the initial energy spread $\sigma _{\varepsilon}$ (see
Fig.\,1). A broad-band laser beam $LB1$ is situated at the orbit region
($\rho _{ph}$, $\rho _{ph} - a$), where $a$ is the laser beam width.
The upper edge of the laser beam must have the form of the flat
boundary and its lower part must have smooth decay of density or width.
In such scheme the laser beam overlap only a part of the ion beam.

In this scheme the ions with the largest amplitudes of betatron
oscillations will interact with the laser beam first. At that
the interaction will take place only at the moment when the ion
deviation caused by the betatron oscillations will have approximately
the amplitude value and the deviation will be directed toward the laser
beam. Immediately after the interaction the position of the ion will be
the same (inside the laser beam $LB1$ and near to its edge) but the
instantaneous orbit will be displaced inward in the direction of the
laser beam. It means that after interaction both the position of the
instantaneous orbit and the amplitude of the betatron oscillations will
be changed at the same value. After every interaction the position of
the instantaneous orbit will be nearer and nearer to the laser beam and
the amplitude of the betatron oscillations will be smaller and smaller
until it will reach zero. At the same time the instantaneous orbit
will reach the edge of the laser beam. Up to that time the
instantaneous orbit will go with some velocity $\dot \rho = d\rho/dt$
in the direction of the laser beam and will not get deeper into the
laser beam.

\vskip -20mm
\special{em:linewidth 0.4pt}
\unitlength 1.00mm
\linethickness{0.4pt}
\begin{picture}(145.67,166.00)
\emline{138.00}{86.00}{1}{18.00}{86.00}{2}
\emline{18.00}{82.00}{3}{138.00}{82.00}{4}
\emline{70.67}{60.00}{5}{71.04}{56.74}{6}
\emline{71.04}{56.74}{7}{71.41}{53.73}{8}
\emline{71.41}{53.73}{9}{71.77}{50.97}{10}
\emline{71.77}{50.97}{11}{72.14}{48.46}{12}
\emline{72.14}{48.46}{13}{72.51}{46.20}{14}
\emline{72.51}{46.20}{15}{72.88}{44.18}{16}
\emline{72.88}{44.18}{17}{73.24}{42.42}{18}
\emline{73.24}{42.42}{19}{73.61}{40.90}{20}
\emline{73.61}{40.90}{21}{73.98}{39.63}{22}
\emline{73.98}{39.63}{23}{74.35}{38.61}{24}
\emline{74.35}{38.61}{25}{74.71}{37.84}{26}
\emline{74.71}{37.84}{27}{75.08}{37.32}{28}
\emline{75.08}{37.32}{29}{75.45}{37.05}{30}
\emline{75.45}{37.05}{31}{75.82}{37.02}{32}
\emline{75.82}{37.02}{33}{76.18}{37.24}{34}
\emline{76.18}{37.24}{35}{76.55}{37.72}{36}
\emline{76.55}{37.72}{37}{76.92}{38.44}{38}
\emline{76.92}{38.44}{39}{77.29}{39.41}{40}
\emline{77.29}{39.41}{41}{77.65}{40.62}{42}
\emline{77.65}{40.62}{43}{78.02}{42.09}{44}
\emline{78.02}{42.09}{45}{78.39}{43.80}{46}
\emline{78.39}{43.80}{47}{78.76}{45.77}{48}
\emline{78.76}{45.77}{49}{79.12}{47.98}{50}
\emline{79.12}{47.98}{51}{79.49}{50.44}{52}
\emline{79.49}{50.44}{53}{79.86}{53.15}{54}
\emline{79.86}{53.15}{55}{80.23}{56.11}{56}
\emline{80.23}{56.11}{57}{80.67}{60.00}{58}
\emline{70.67}{60.00}{59}{80.67}{60.00}{60}
\emline{50.67}{86.00}{61}{52.30}{82.63}{62}
\emline{52.30}{82.63}{63}{53.93}{79.50}{64}
\emline{53.93}{79.50}{65}{55.56}{76.58}{66}
\emline{55.56}{76.58}{67}{57.19}{73.90}{68}
\emline{57.19}{73.90}{69}{58.82}{71.44}{70}
\emline{58.82}{71.44}{71}{60.45}{69.21}{72}
\emline{60.45}{69.21}{73}{62.09}{67.21}{74}
\emline{62.09}{67.21}{75}{63.72}{65.43}{76}
\emline{63.72}{65.43}{77}{65.35}{63.88}{78}
\emline{65.35}{63.88}{79}{66.98}{62.56}{80}
\emline{66.98}{62.56}{81}{68.61}{61.46}{82}
\emline{68.61}{61.46}{83}{70.24}{60.59}{84}
\emline{70.24}{60.59}{85}{71.87}{59.95}{86}
\emline{71.87}{59.95}{87}{73.50}{59.54}{88}
\emline{73.50}{59.54}{89}{75.13}{59.35}{90}
\emline{75.13}{59.35}{91}{76.76}{59.39}{92}
\emline{76.76}{59.39}{93}{78.39}{59.65}{94}
\emline{78.39}{59.65}{95}{80.02}{60.14}{96}
\emline{80.02}{60.14}{97}{81.66}{60.86}{98}
\emline{81.66}{60.86}{99}{83.29}{61.81}{100}
\emline{83.29}{61.81}{101}{84.92}{62.98}{102}
\emline{84.92}{62.98}{103}{86.55}{64.38}{104}
\emline{86.55}{64.38}{105}{88.18}{66.01}{106}
\emline{88.18}{66.01}{107}{89.81}{67.86}{108}
\emline{89.81}{67.86}{109}{91.44}{69.95}{110}
\emline{91.44}{69.95}{111}{93.07}{72.25}{112}
\emline{93.07}{72.25}{113}{94.70}{74.79}{114}
\emline{94.70}{74.79}{115}{96.33}{77.55}{116}
\emline{96.33}{77.55}{117}{97.96}{80.54}{118}
\emline{97.96}{80.54}{119}{100.67}{86.00}{120}
\emline{50.67}{82.00}{121}{52.40}{78.93}{122}
\emline{52.40}{78.93}{123}{54.13}{76.08}{124}
\emline{54.13}{76.08}{125}{55.86}{73.44}{126}
\emline{55.86}{73.44}{127}{57.59}{71.03}{128}
\emline{57.59}{71.03}{129}{59.32}{68.84}{130}
\emline{59.32}{68.84}{131}{61.05}{66.87}{132}
\emline{61.05}{66.87}{133}{62.77}{65.12}{134}
\emline{62.77}{65.12}{135}{64.50}{63.59}{136}
\emline{64.50}{63.59}{137}{66.23}{62.28}{138}
\emline{66.23}{62.28}{139}{67.96}{61.19}{140}
\emline{67.96}{61.19}{141}{69.69}{60.32}{142}
\emline{69.69}{60.32}{143}{71.42}{59.66}{144}
\emline{71.42}{59.66}{145}{73.15}{59.23}{146}
\emline{73.15}{59.23}{147}{74.88}{59.02}{148}
\emline{74.88}{59.02}{149}{76.61}{59.03}{150}
\emline{76.61}{59.03}{151}{78.34}{59.26}{152}
\emline{78.34}{59.26}{153}{80.07}{59.71}{154}
\emline{80.07}{59.71}{155}{81.80}{60.38}{156}
\emline{81.80}{60.38}{157}{83.53}{61.27}{158}
\emline{83.53}{61.27}{159}{85.26}{62.38}{160}
\emline{85.26}{62.38}{161}{86.98}{63.71}{162}
\emline{86.98}{63.71}{163}{88.71}{65.26}{164}
\emline{88.71}{65.26}{165}{90.44}{67.03}{166}
\emline{90.44}{67.03}{167}{92.17}{69.02}{168}
\emline{92.17}{69.02}{169}{93.90}{71.23}{170}
\emline{93.90}{71.23}{171}{95.63}{73.66}{172}
\emline{95.63}{73.66}{173}{97.36}{76.31}{174}
\emline{97.36}{76.31}{175}{99.09}{79.18}{176}
\emline{99.09}{79.18}{177}{100.67}{82.00}{178}
\emline{50.67}{78.00}{179}{52.53}{75.28}{180}
\emline{52.53}{75.28}{181}{54.39}{72.77}{182}
\emline{54.39}{72.77}{183}{56.24}{70.47}{184}
\emline{56.24}{70.47}{185}{58.10}{68.38}{186}
\emline{58.10}{68.38}{187}{59.96}{66.50}{188}
\emline{59.96}{66.50}{189}{61.82}{64.83}{190}
\emline{61.82}{64.83}{191}{63.68}{63.37}{192}
\emline{63.68}{63.37}{193}{65.54}{62.12}{194}
\emline{65.54}{62.12}{195}{67.39}{61.08}{196}
\emline{67.39}{61.08}{197}{69.25}{60.25}{198}
\emline{69.25}{60.25}{199}{71.11}{59.63}{200}
\emline{71.11}{59.63}{201}{72.97}{59.22}{202}
\emline{72.97}{59.22}{203}{74.83}{59.02}{204}
\emline{74.83}{59.02}{205}{76.68}{59.03}{206}
\emline{76.68}{59.03}{207}{78.54}{59.25}{208}
\emline{78.54}{59.25}{209}{80.40}{59.68}{210}
\emline{80.40}{59.68}{211}{82.26}{60.32}{212}
\emline{82.26}{60.32}{213}{84.12}{61.17}{214}
\emline{84.12}{61.17}{215}{85.98}{62.23}{216}
\emline{85.98}{62.23}{217}{87.83}{63.50}{218}
\emline{87.83}{63.50}{219}{89.69}{64.98}{220}
\emline{89.69}{64.98}{221}{91.55}{66.67}{222}
\emline{91.55}{66.67}{223}{93.41}{68.57}{224}
\emline{93.41}{68.57}{225}{95.27}{70.67}{226}
\emline{95.27}{70.67}{227}{97.13}{72.99}{228}
\emline{97.13}{72.99}{229}{98.98}{75.52}{230}
\emline{98.98}{75.52}{231}{100.67}{78.00}{232}
\emline{25.67}{113.00}{233}{27.03}{113.05}{234}
\emline{27.03}{113.05}{235}{28.41}{112.91}{236}
\emline{28.41}{112.91}{237}{29.79}{112.56}{238}
\emline{29.79}{112.56}{239}{31.19}{112.01}{240}
\emline{31.19}{112.01}{241}{32.60}{111.27}{242}
\emline{32.60}{111.27}{243}{34.01}{110.32}{244}
\emline{34.01}{110.32}{245}{35.44}{109.17}{246}
\emline{35.44}{109.17}{247}{36.88}{107.82}{248}
\emline{36.88}{107.82}{249}{38.34}{106.27}{250}
\emline{38.34}{106.27}{251}{39.80}{104.52}{252}
\emline{39.80}{104.52}{253}{41.28}{102.57}{254}
\emline{41.28}{102.57}{255}{42.76}{100.42}{256}
\emline{42.76}{100.42}{257}{44.26}{98.07}{258}
\emline{44.26}{98.07}{259}{45.77}{95.51}{260}
\emline{45.77}{95.51}{261}{47.29}{92.76}{262}
\emline{47.29}{92.76}{263}{48.82}{89.81}{264}
\emline{48.82}{89.81}{265}{50.67}{86.00}{266}
\emline{25.67}{105.00}{267}{26.97}{104.98}{268}
\emline{26.97}{104.98}{269}{28.31}{104.75}{270}
\emline{28.31}{104.75}{271}{29.71}{104.32}{272}
\emline{29.71}{104.32}{273}{31.15}{103.67}{274}
\emline{31.15}{103.67}{275}{32.64}{102.81}{276}
\emline{32.64}{102.81}{277}{34.19}{101.74}{278}
\emline{34.19}{101.74}{279}{35.78}{100.45}{280}
\emline{35.78}{100.45}{281}{37.42}{98.96}{282}
\emline{37.42}{98.96}{283}{39.11}{97.26}{284}
\emline{39.11}{97.26}{285}{40.84}{95.35}{286}
\emline{40.84}{95.35}{287}{42.63}{93.22}{288}
\emline{42.63}{93.22}{289}{44.47}{90.89}{290}
\emline{44.47}{90.89}{291}{46.35}{88.34}{292}
\emline{46.35}{88.34}{293}{48.28}{85.59}{294}
\emline{48.28}{85.59}{295}{50.67}{82.00}{296}
\emline{25.67}{97.00}{297}{26.90}{97.08}{298}
\emline{26.90}{97.08}{299}{28.22}{96.94}{300}
\emline{28.22}{96.94}{301}{29.62}{96.58}{302}
\emline{29.62}{96.58}{303}{31.11}{95.99}{304}
\emline{31.11}{95.99}{305}{32.68}{95.19}{306}
\emline{32.68}{95.19}{307}{34.34}{94.16}{308}
\emline{34.34}{94.16}{309}{36.09}{92.92}{310}
\emline{36.09}{92.92}{311}{37.91}{91.45}{312}
\emline{37.91}{91.45}{313}{39.83}{89.76}{314}
\emline{39.83}{89.76}{315}{41.83}{87.85}{316}
\emline{41.83}{87.85}{317}{43.91}{85.72}{318}
\emline{43.91}{85.72}{319}{46.08}{83.37}{320}
\emline{46.08}{83.37}{321}{48.33}{80.79}{322}
\emline{48.33}{80.79}{323}{50.67}{78.00}{324}
\emline{100.67}{86.00}{325}{102.32}{89.27}{326}
\emline{102.32}{89.27}{327}{103.95}{92.33}{328}
\emline{103.95}{92.33}{329}{105.56}{95.17}{330}
\emline{105.56}{95.17}{331}{107.14}{97.81}{332}
\emline{107.14}{97.81}{333}{108.69}{100.24}{334}
\emline{108.69}{100.24}{335}{110.22}{102.46}{336}
\emline{110.22}{102.46}{337}{111.73}{104.47}{338}
\emline{111.73}{104.47}{339}{113.21}{106.27}{340}
\emline{113.21}{106.27}{341}{114.66}{107.86}{342}
\emline{114.66}{107.86}{343}{116.10}{109.24}{344}
\emline{116.10}{109.24}{345}{117.50}{110.41}{346}
\emline{117.50}{110.41}{347}{118.88}{111.38}{348}
\emline{118.88}{111.38}{349}{120.24}{112.13}{350}
\emline{120.24}{112.13}{351}{121.57}{112.67}{352}
\emline{121.57}{112.67}{353}{122.88}{113.00}{354}
\emline{122.88}{113.00}{355}{124.16}{113.13}{356}
\emline{124.16}{113.13}{357}{125.67}{113.00}{358}
\emline{125.67}{105.00}{359}{124.41}{104.98}{360}
\emline{124.41}{104.98}{361}{123.10}{104.75}{362}
\emline{123.10}{104.75}{363}{121.74}{104.32}{364}
\emline{121.74}{104.32}{365}{120.32}{103.67}{366}
\emline{120.32}{103.67}{367}{118.85}{102.81}{368}
\emline{118.85}{102.81}{369}{117.32}{101.74}{370}
\emline{117.32}{101.74}{371}{115.73}{100.45}{372}
\emline{115.73}{100.45}{373}{114.09}{98.96}{374}
\emline{114.09}{98.96}{375}{112.40}{97.26}{376}
\emline{112.40}{97.26}{377}{110.65}{95.35}{378}
\emline{110.65}{95.35}{379}{108.85}{93.22}{380}
\emline{108.85}{93.22}{381}{106.99}{90.89}{382}
\emline{106.99}{90.89}{383}{105.07}{88.34}{384}
\emline{105.07}{88.34}{385}{103.10}{85.59}{386}
\emline{103.10}{85.59}{387}{100.67}{82.00}{388}
\emline{125.67}{97.00}{389}{124.30}{97.08}{390}
\emline{124.30}{97.08}{391}{122.87}{96.94}{392}
\emline{122.87}{96.94}{393}{121.38}{96.58}{394}
\emline{121.38}{96.58}{395}{119.82}{95.99}{396}
\emline{119.82}{95.99}{397}{118.20}{95.19}{398}
\emline{118.20}{95.19}{399}{116.51}{94.16}{400}
\emline{116.51}{94.16}{401}{114.76}{92.92}{402}
\emline{114.76}{92.92}{403}{112.94}{91.45}{404}
\emline{112.94}{91.45}{405}{111.05}{89.76}{406}
\emline{111.05}{89.76}{407}{109.11}{87.85}{408}
\emline{109.11}{87.85}{409}{107.09}{85.72}{410}
\emline{107.09}{85.72}{411}{105.02}{83.37}{412}
\emline{105.02}{83.37}{413}{102.88}{80.79}{414}
\emline{102.88}{80.79}{415}{100.67}{78.00}{416}
\emline{50.67}{66.00}{417}{53.02}{64.74}{418}
\emline{53.02}{64.74}{419}{55.37}{63.61}{420}
\emline{55.37}{63.61}{421}{57.73}{62.61}{422}
\emline{57.73}{62.61}{423}{60.08}{61.72}{424}
\emline{60.08}{61.72}{425}{62.43}{60.96}{426}
\emline{62.43}{60.96}{427}{64.78}{60.33}{428}
\emline{64.78}{60.33}{429}{67.13}{59.82}{430}
\emline{67.13}{59.82}{431}{69.48}{59.43}{432}
\emline{69.48}{59.43}{433}{71.84}{59.16}{434}
\emline{71.84}{59.16}{435}{74.19}{59.02}{436}
\emline{74.19}{59.02}{437}{76.54}{59.01}{438}
\emline{76.54}{59.01}{439}{78.89}{59.12}{440}
\emline{78.89}{59.12}{441}{81.24}{59.35}{442}
\emline{81.24}{59.35}{443}{83.59}{59.70}{444}
\emline{83.59}{59.70}{445}{85.95}{60.18}{446}
\emline{85.95}{60.18}{447}{88.30}{60.79}{448}
\emline{88.30}{60.79}{449}{90.65}{61.51}{450}
\emline{90.65}{61.51}{451}{93.00}{62.36}{452}
\emline{93.00}{62.36}{453}{95.35}{63.34}{454}
\emline{95.35}{63.34}{455}{97.70}{64.44}{456}
\emline{97.70}{64.44}{457}{100.67}{66.00}{458}
\emline{50.67}{62.00}{459}{53.16}{61.43}{460}
\emline{53.16}{61.43}{461}{55.64}{60.92}{462}
\emline{55.64}{60.92}{463}{58.13}{60.48}{464}
\emline{58.13}{60.48}{465}{60.62}{60.09}{466}
\emline{60.62}{60.09}{467}{63.11}{59.76}{468}
\emline{63.11}{59.76}{469}{65.59}{59.49}{470}
\emline{65.59}{59.49}{471}{68.08}{59.28}{472}
\emline{68.08}{59.28}{473}{70.57}{59.12}{474}
\emline{70.57}{59.12}{475}{73.06}{59.03}{476}
\emline{73.06}{59.03}{477}{78.03}{59.03}{478}
\emline{78.03}{59.03}{479}{80.52}{59.11}{480}
\emline{80.52}{59.11}{481}{83.01}{59.26}{482}
\emline{83.01}{59.26}{483}{85.49}{59.46}{484}
\emline{85.49}{59.46}{485}{87.98}{59.73}{486}
\emline{87.98}{59.73}{487}{90.47}{60.05}{488}
\emline{90.47}{60.05}{489}{92.96}{60.43}{490}
\emline{92.96}{60.43}{491}{95.44}{60.88}{492}
\emline{95.44}{60.88}{493}{97.93}{61.38}{494}
\emline{97.93}{61.38}{495}{100.67}{62.00}{496}
\emline{50.67}{62.00}{497}{48.26}{62.57}{498}
\emline{48.26}{62.57}{499}{45.83}{63.08}{500}
\emline{45.83}{63.08}{501}{43.38}{63.53}{502}
\emline{43.38}{63.53}{503}{40.91}{63.92}{504}
\emline{40.91}{63.92}{505}{38.42}{64.25}{506}
\emline{38.42}{64.25}{507}{35.91}{64.52}{508}
\emline{35.91}{64.52}{509}{33.38}{64.73}{510}
\emline{33.38}{64.73}{511}{30.83}{64.88}{512}
\emline{30.83}{64.88}{513}{28.26}{64.97}{514}
\emline{28.26}{64.97}{515}{25.67}{65.00}{516}
\emline{25.67}{73.00}{517}{27.85}{72.94}{518}
\emline{27.85}{72.94}{519}{30.05}{72.76}{520}
\emline{30.05}{72.76}{521}{32.28}{72.46}{522}
\emline{32.28}{72.46}{523}{34.54}{72.04}{524}
\emline{34.54}{72.04}{525}{36.83}{71.50}{526}
\emline{36.83}{71.50}{527}{39.15}{70.84}{528}
\emline{39.15}{70.84}{529}{41.50}{70.06}{530}
\emline{41.50}{70.06}{531}{43.87}{69.16}{532}
\emline{43.87}{69.16}{533}{46.27}{68.14}{534}
\emline{46.27}{68.14}{535}{50.67}{66.00}{536}
\emline{100.67}{66.00}{537}{103.16}{67.16}{538}
\emline{103.16}{67.16}{539}{105.63}{68.23}{540}
\emline{105.63}{68.23}{541}{108.09}{69.18}{542}
\emline{108.09}{69.18}{543}{110.52}{70.03}{544}
\emline{110.52}{70.03}{545}{112.94}{70.78}{546}
\emline{112.94}{70.78}{547}{115.34}{71.42}{548}
\emline{115.34}{71.42}{549}{117.72}{71.96}{550}
\emline{117.72}{71.96}{551}{120.08}{72.39}{552}
\emline{120.08}{72.39}{553}{122.42}{72.72}{554}
\emline{122.42}{72.72}{555}{125.67}{73.00}{556}
\emline{100.67}{62.00}{557}{103.38}{62.57}{558}
\emline{103.38}{62.57}{559}{106.04}{63.08}{560}
\emline{106.04}{63.08}{561}{108.66}{63.53}{562}
\emline{108.66}{63.53}{563}{111.23}{63.92}{564}
\emline{111.23}{63.92}{565}{113.76}{64.25}{566}
\emline{113.76}{64.25}{567}{116.23}{64.52}{568}
\emline{116.23}{64.52}{569}{118.66}{64.73}{570}
\emline{118.66}{64.73}{571}{121.04}{64.88}{572}
\emline{121.04}{64.88}{573}{123.38}{64.97}{574}
\emline{123.38}{64.97}{575}{125.67}{65.00}{576}
\emline{25.67}{58.00}{577}{125.67}{58.00}{578}
\put(16.00,86.00){\makebox(0,0)[cc]{$1$}}
\put(16.00,82.00){\makebox(0,0)[cc]{$2$}}
\put(20.67,113.00){\makebox(0,0)[cc]{$1$}}
\put(20.67,105.00){\makebox(0,0)[cc]{$2$}}
\put(20.67,97.00){\makebox(0,0)[cc]{$3$}}
\put(20.67,73.00){\makebox(0,0)[cc]{$6$}}
\put(20.67,65.00){\makebox(0,0)[cc]{$7$}}
\put(20.67,58.00){\makebox(0,0)[cc]{$8$}}
\emline{25.67}{54.00}{579}{125.67}{54.00}{580}
\emline{125.67}{50.00}{581}{25.67}{50.00}{582}
\emline{25.67}{46.00}{583}{125.67}{46.00}{584}
\put(20.67,54.00){\makebox(0,0)[cc]{$9$}}
\put(20.67,50.00){\makebox(0,0)[cc]{$10$}}
\put(20.67,46.00){\makebox(0,0)[cc]{$11$}}
\emline{70.67}{120.00}{585}{80.67}{120.00}{586}
\emline{80.67}{120.00}{587}{80.30}{123.26}{588}
\emline{80.30}{123.26}{589}{79.93}{126.27}{590}
\emline{79.93}{126.27}{591}{79.57}{129.03}{592}
\emline{79.57}{129.03}{593}{79.20}{131.54}{594}
\emline{79.20}{131.54}{595}{78.83}{133.80}{596}
\emline{78.83}{133.80}{597}{78.46}{135.82}{598}
\emline{78.46}{135.82}{599}{78.10}{137.58}{600}
\emline{78.10}{137.58}{601}{77.73}{139.10}{602}
\emline{77.73}{139.10}{603}{77.36}{140.37}{604}
\emline{77.36}{140.37}{605}{76.99}{141.39}{606}
\emline{76.99}{141.39}{607}{76.63}{142.16}{608}
\emline{76.63}{142.16}{609}{76.26}{142.68}{610}
\emline{76.26}{142.68}{611}{75.89}{142.95}{612}
\emline{75.89}{142.95}{613}{75.52}{142.98}{614}
\emline{75.52}{142.98}{615}{75.16}{142.76}{616}
\emline{75.16}{142.76}{617}{74.79}{142.28}{618}
\emline{74.79}{142.28}{619}{74.42}{141.56}{620}
\emline{74.42}{141.56}{621}{74.05}{140.59}{622}
\emline{74.05}{140.59}{623}{73.69}{139.38}{624}
\emline{73.69}{139.38}{625}{73.32}{137.91}{626}
\emline{73.32}{137.91}{627}{72.95}{136.20}{628}
\emline{72.95}{136.20}{629}{72.58}{134.23}{630}
\emline{72.58}{134.23}{631}{72.22}{132.02}{632}
\emline{72.22}{132.02}{633}{71.85}{129.56}{634}
\emline{71.85}{129.56}{635}{71.48}{126.85}{636}
\emline{71.48}{126.85}{637}{71.11}{123.89}{638}
\emline{71.11}{123.89}{639}{70.67}{120.00}{640}
\emline{75.67}{121.67}{641}{89.67}{133.00}{642}
\emline{89.67}{133.00}{643}{99.00}{133.00}{644}
\emline{75.67}{43.00}{645}{90.00}{31.00}{646}
\emline{90.00}{31.00}{647}{99.00}{31.00}{648}
\put(94.34,33.33){\makebox(0,0)[cc]{$LB1$}}
\put(94.34,135.00){\makebox(0,0)[cc]{$LB2$}}
\put(140.67,86.00){\makebox(0,0)[cc]{$1$}}
\put(140.67,82.00){\makebox(0,0)[cc]{$2$}}
\put(145.67,90.00){\vector(1,0){0.2}}
\emline{8.34}{90.00}{649}{145.67}{90.00}{650}
\put(145.67,86.33){\makebox(0,0)[cc]{$y$}}
\put(75.67,70.00){\vector(0,1){0.2}}
\emline{75.67}{60.00}{651}{75.67}{70.00}{652}
\put(79.34,68.67){\makebox(0,0)[cc]{$\vec v_1$}}
\put(75.67,110.00){\vector(0,-1){0.2}}
\emline{75.67}{120.00}{653}{75.67}{110.00}{654}
\put(79.34,112.33){\makebox(0,0)[cc]{$\vec v_2$}}
\put(75.67,20.67){\makebox(0,0)[cc]
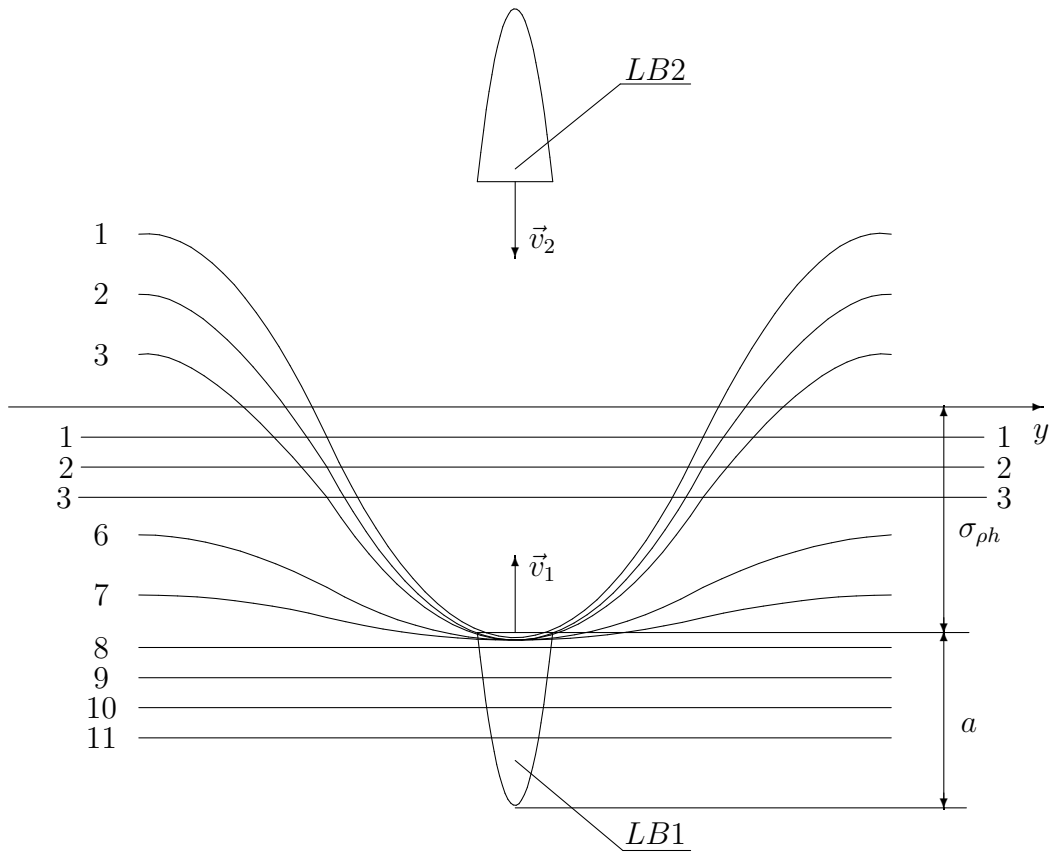
{\small Fig.1: The scheme of the two-dimensional ion cooling. The axis $y$ is the equilibrium}}
\put(77.00,15.67){\makebox(0,0)[cc]
{\small orbit of the storage ring, 1-1, 2-2 ... the location of the instantaneous ion orbit }}
\put(76.33,10.67){\makebox(0,0)[cc]
{\small after 0,1,2 ... events of the ion energy loss, $LB1$ and $LB2$ are the laser beams }}
\put(74.67,5.67){\makebox(0,0)[cc]
{\small moving by turns with the velocity $v_{1,2}$ from outside to the equilibrium orbit. }}
\emline{136.00}{60.00}{655}{80.33}{60.00}{656}
\put(132.67,60.00){\vector(0,-1){0.2}}
\emline{132.67}{62.00}{657}{132.67}{60.00}{658}
\emline{135.67}{36.67}{659}{75.67}{36.67}{660}
\put(132.67,60.00){\vector(0,1){0.2}}
\emline{132.67}{36.67}{661}{132.67}{60.00}{662}
\put(132.67,36.67){\vector(0,-1){0.2}}
\emline{132.67}{38.67}{663}{132.67}{36.67}{664}
\put(136.00,47.67){\makebox(0,0)[cc]{$a$}}
\put(137.50,73.00){\makebox(0,0)[cc]{$\sigma _{\rho h}$}}
\emline{17.67}{78.00}{665}{138.33}{78.00}{666}
\put(140.67,78.00){\makebox(0,0)[cc]{$3$}}
\put(15.67,78.00){\makebox(0,0)[cc]{$3$}}
\put(132.67,90.00){\vector(0,1){0.2}}
\emline{132.67}{60.00}{667}{132.67}{90.00}{668}
\end{picture}

\vskip 5mm
At the moment when the instantaneous orbit will enter the laser beam
the amplitude of the ion betatron oscillations will be
small. After the instantaneous orbit will enter the laser beam then it
will continue its movement but now it will move through the laser
beam. Now the amplitude will be increasing proportionally to the
root-mean-square of the number of the interactions of the ion with the
laser beam that is slowly then the decreasing of the amplitude in the
previous case (proportionally to the number of the interactions). This
is because of the ion will interact with the laser beam both with
positive and negative deviation from the instantaneous orbit.

The ion beam has a set of amplitudes of betatron oscillations and
instantaneous orbits. To cool all ions of the beam we must move the
laser beam radial position with some velocity $v_1$ in the direction of
the ion beam or move the instantaneous orbits by kick, phase
displacement or eddy electric fields in the direction of the laser
beam. This velocity must be much less then the velocity of the
instantaneous orbit $\dot \rho$. In this case the amplitudes
of the betatron oscillations will be in time to decrease theirs
amplitudes to small values before they enter the laser beam. After all
ions of the ion beam will be in the state of interaction with the
laser beam then the laser beam must be stopped. After the stop the
ions will continue to move to the tapered edge of the laser beam with
the decreasing velocity and will be gathered at this edge. The longer
we will wait the less the energy spread will be obtained.

As the thickness of the laser beam will be the less the nearer the ion
orbit to the remote edge of the laser beam then the damping time will
be too high if we will wait for a small energy spread of the cooled ion
beam. To shorten this damping time we can use the second laser beam
from the opposite side of the ion beam (see Fig.1). At this stage the
ion beam will have small amplitudes of betatron oscillations. The
second laser beam radial position must be moved to the ion beam with
the velocity slightly higher then the average velocity of movement of
the instantaneous ion orbit which is determined by the intensity of the
laser beam. In this case the ion orbits will be in time to enter into
the laser beam at a distance higher then the residual amplitude of ion
betatron oscillations and hence the amplitude of betatron oscillations
at this stage will be increased slowly with the number of interactions
(according to the root-mean-square law). High energy ions first of all
and then ions with smaller energies will interact with the laser beam.
This process will last until laser beam will reach the instantaneous
orbit with the least energy and ions of this orbit will be cooled. Such
a way we can repeat the cooling process and to reach the energy spread
of the ion beam

        \begin{equation}
        {\Delta \varepsilon _{i} \over \varepsilon _{i}} \simeq \pm
        {\varepsilon _{int} \over Mc^2 \gamma},  \end{equation} 
where $\varepsilon _{i} = Mc^2\gamma $ is the ion energy, $M$ ion rest
mass, $\varepsilon _{int} $ the average energy of the ion loss per one
event of interaction, $\gamma $ the relativistic factor of the ion.

The second laser beam can be monochromatic one with the sweeping
frequency and overlap the ion beam as a whole. In this case the
ordinary one-dimensional laser cooling will take place at the second
stage of the laser cooling.

The two-step scheme will work in the case when the RF system of the
storage ring is switched on as well. Some peculiarities will be in this
case.

The number of interactions of one ion with the laser beam per one
collision of the ion with a laser beam

        \begin{equation}
        \Delta N _{int} = (1 + \beta){I _L\over \hbar \omega
        _L}{l _L\over c} \sigma,  \end{equation} 
where $\beta =v/c$, $v$ is the ion velocity, $I _L$ the laser intensity
(power per unit area), $\hbar \omega _L$ the energy of the laser
photon, $l _L$ the length of the interaction region of the laser and
ion beams, $\sigma $ the cross section of the process of the laser
photon-ion interaction.

The damping time of the ion beam in this scheme is $\tau = \tau _s +
\tau _b$, where

        \begin{equation}
        \tau _{s} = {\sigma _{\varepsilon}\over k_1 f \Delta N _{int}
        \varepsilon _{int}},
\hskip 5mm
        \tau _{b} = {\sigma _{\varepsilon b}\over k_2 f \Delta N
        _{int} \varepsilon _{int}},         \end{equation} 
$\tau _s$ is the ion cooling time for the longitudinal (energy -
length) space, $\tau _b$ the ion cooling time for the radial betatron
oscillations, $f$ the frequency of the ion beam revolution in the
storage ring, $\sigma _{b\varepsilon} = \sigma _b\varepsilon/\alpha
\overline R$ the energy interval corresponding to one for the
instantaneous orbits distributed through the interval of radii $\sigma
_b$, $\overline R$ the average radius of the storage ring, $k_1$, $k_2$
are the ratio of the velocity of movement of the first and second laser
beam to the average velocity of movement of the ion instantaneous orbit
caused by the interaction of the ions with the laser beam respectively,
$\alpha $ the momentum compaction function.

We now considere two examples of the two-dimensional ion cooling. First
the hydrogenlike $Pb$ ion beam in the CERN LHC, for which the relevant
parameters are $2\pi \overline R = 27$ km, $\gamma = 3000$, $Mc^2\gamma
= 575$ TeV, $Z = 82$, $\sigma _x = 1.2\cdot 10^{-3}$, $\Delta
\gamma/\gamma = 2\cdot 10^{-4}$ ($\sigma _{\varepsilon } = 1.15\cdot
10^{11}$ eV). Ion beam cooling occurs through the backward Rayleigh 
scattering of the laser photons. For the bandwidth of the laser beam 
$\Delta \omega _L/\omega _L = 10 ^{-4}$, the transition between the 
$1S$ ground state and the $2P$ excited state of a hydrogenlike $Pb$ 
ion, we obtain from the well-known formulas the value of the resonant 
transition energy $\hbar \omega ^{*} = 68.7$ keV, the cross section 
$\sigma = 6.58 \cdot 10^{-18 }$ cm$^2$, average energy of the scattered 
photon $<\hbar \omega ^s> = \varepsilon _{int} = 0.2$ GeV, and the 
required laser wavelength $\lambda _L = 1080 \AA$ \cite{prl}. The laser 
cross section is larger then that of the ion beam as long as the 
Rayleigh length $z_R = \pi \sigma _x^2/\lambda _L \geq 0.42$ cm. We 
will choose the value $l_L = 2 z_R = 15$ cm for the rms transverse 
laser beam size at its waist $\sigma _L = 5.8 \cdot 10^{-3}$ cm and the 
power of the laser beam is $P_L = 400$ W. In this case $I_L = 4.94$ 
MW/cm$^2$, $\hbar \omega _L = 11.49$ eV, $f = 1.11 \cdot 10 ^{4}$ Hz, 
$\Delta N_{int} = 1.68 \cdot 10 ^{-2}$, the damping time $\tau _s|_{k_1 
= 10} \simeq 3.06$ sec. The damping time for the betatron oscillations 
$\tau _b|_{k_2 = 1.1} \simeq 0.28$ sec when the value $\sigma 
_{\varepsilon b} = \sigma _{\varepsilon }$. The limiting relative 
energy spread (1) in this case $\Delta \varepsilon _i/\varepsilon _i 
\simeq 3.5 \cdot 10^{-7}$. A metal screen can be used to produce the 
laser beam with sharp edge.

In the second example the ion cooling of fully stripped $Pb$ ion beam
of LHC occurs through the electron-positron pair production  \cite{bw},
\cite{bkw}. The parameters of the ion beam and the cross section of
the laser beam are those of the previous example. In this case the
threshold energy of the laser photon beam $\hbar \omega _{L, thr} = 335
eV$. We choose the photon energy $\hbar \omega _L = 670 eV$ and the
power $P_L = 10^6$ W.  In this case $I_L = 1.23 \cdot 10^{10}$
W/cm$^2$, $\lambda _L = 18.5$ $\AA$, $z_R = 57$ cm, $l_L = 115$ cm,
$\sigma = 8.1 \cdot 10^{-24}$ cm$^2$, $\varepsilon _{int} \simeq 2mc^2
\simeq 3.06$ GeV \cite{bkw}, $\Delta N _{int} = 6.84 \cdot 10^{-6}$,
$\tau _s|_{k_1 = 10} = 4.95 \cdot 10^3$ sec $\simeq 1.38$ h, $\tau
_b|_{k_2 = 1.1} = 4.5 \cdot 10^2$ sec. The average power $\overline P_L
= 10^6$ W in $1 \AA$ region is not real now.  But if all particles of 
the ion beam will be gathered in one bunch of the length $l_{ion} = l 
_L/2 = z_R$ and the laser beam will consist of the wavepackets of the 
length $l_L = 2 z_R$ which follow with the frequency $f$, then the 
average laser power can be decreased $2\pi \overline R/l_L = 2.35 \cdot 
10^{4}$ times that is to the value $\overline P = 43$ W at the same 
damping time $\tau _s|_{k_1 = 10} \simeq 1.38$ h. The limiting relative 
energy spread (1) in this case $\Delta \varepsilon _i/\varepsilon _i 
\simeq 2.7 \cdot 10^{-6}$.

We can get the part of the LHC trajectory with the average value of the
$\beta $-function $\overline \beta = 62$ m \cite{at/94} to work with
higher dimensions of the ion beam. Such interaction can take place in
the banding magnet of the LHC and the laser beam with sharp edge can
fall at some small angle to the plane of the ion orbit in order to the
interaction took place at short distance $l_L$ and the boundary of the
laser beam was flat through the interaction region.

The examples show that the suggested version of the two-dimensional
cooling is more efficient then the three-dimensional radiative laser
cooling scheme for the case of the resonant Rayleigh backscattering
scheme \cite{prl}. The ion cooling through electron-positron pair
production is possible in principle in this scheme. Some problems can
arise for the heavy ion cooling in the last case because of strong
Z-dependence of the probability of the electron capture by ion from the
produced pair \cite{at/94}. The scheme under consideration based on the
backward Compton scattering can be used in the case of the electron
cooling as well.

One of the authors (E.G.) thanks Dr.Karlheinz Schindl who paid attention
on strong Z-dependence of the probability of the electron capture by ion
from the produced pair.

Support of this work by Russian Fund of the Foundation Research Grant
No 96-02-18167 is gratefully acknowledged.

       
\end{document}